# Critical current density and trapped field in HTS with asymmetric magnetization loops

D Gokhfeld

Kirensky Institute of Physics SB RAS, Akademgorodok 50/38, Krasnoyarsk 66036, Russian Federation

E-mail: gokhfeld@iph.krasn.ru

**Abstract**. Applications of the extended critical state model are considered. The trapped magnetic field, the penetration field and the field dependence of the critical current density are analysed. The critical current density and the trapped field in superconducting grains depend on the grain size. Asymmetry of the hysteresis curves relative to the $M = 0$ axis is related to the scale of the current circulation.

## 1. Introduction

The extended critical state model (ECSM) [1-4] is recent modification of the critical state model describing magnetization loops of type II superconductors. ECSM suggests that the asymmetry of the magnetization loop relative to the $M = 0$ axis is due to the equilibrium magnetization of the surface layer. The total magnetization of a sample is the sum of the equilibrium magnetization of the surface layer and the nonequilibrium magnetization of the inner volume. In the surface layer, the vortices are not pinned due to the interaction with the screening currents and the surface [5, 6].

ECSM combined with the computation of the field distribution into the sample [2-4] gives new facilities for analysis of magnetic characteristics of bulk superconductors. This model is applicable to describe magnetization loops of different superconducting materials [2, 7-13]. Estimations with using ECSM are useful for constructing of promising application devises (trapped field magnets, magnetic shielding, levitation etc.).

In this work some application aspects of ECSM are considered. The trapped field in superconducting samples is described in subsection 3.1. Parameters estimated from fitting of experimental magnetization loops result in phenomenological field dependence of the critical current density (subsection 3.2) and formula for the full penetration field (subsection 3.3). Determining of the screening current scale is discussed in subsection 3.4.

## 2. Extended critical state model

A cylindrical sample with the length significantly larger than the size of the base is under consideration. The demagnetization factor of such sample can be taken as zero. A thin plate oriented along the magnetic field can be considered in similar manner [4]. The magnetization $M$ of a type II superconductor is defined as the diamagnetic response $-H$ plus the averaged magnetic field inside the sample, here $H$ is the external magnetic field. For an infinitely long cylindrical sample with radius $R$, which is coaxial with the external magnetic field, the magnetization is determined by the following expression

$$M(H) = -H + \frac{2}{\mu_0 R^2} \int_0^R r B(r) dr,$$

(1)

where $r$ is the distance from the cylinder axis, $B$ is the local magnetic field in the sample, $\mu_0$ is the magnetic constant. The distribution of the magnetic field $B(r)$ (the flux density) inside the sample is calculated from Ampere's circuital law $dB/dr = \mu_0 j_c(B)$ following approach [14], here $j_c$ is the critical current density.

The dependence $M(H)$ is computed with the calculated distributions $B(r)$ for all $H$. It takes into account that the field distribution in the surface layer does not depend on the magnetization prehistory. The surface layer with the equilibrium magnetization has the depth $l_s$ about the depth of magnetic field penetration $\lambda$ [5, 6]. The asymmetry of the magnetization loops is determined by the fraction $l_s/R$.

The main steps of the parameterization and the calculation of the magnetization loop of a superconductor in ECSM are enumerated in work [4]. Parameters are chosen from the best fit of the calculated dependences to experimental magnetization loops. The calculated $M(H)$ curves for different values of $l_s/R$ are presented on figure 1. To compute these loops the irreversibility field $H_{irr}$ is established to be much smaller than the critical field $H_{c2}$. Such case is relevant for high temperature superconductors (HTS) [4]. As the value of $l_s/R$ growths from 0 to 1 the asymmetry of the loops increases. The fully reversible curve for $l_s/R = 1$ (maximal asymmetry) coincides with the virgin magnetization branches of all curves. The critical current density dependencies on the magnetic field $j_c(H)$ extracted from the $M(H)$ loops with using ECSM and the Bean model are presented on figure 2 (see subsection 3.2).

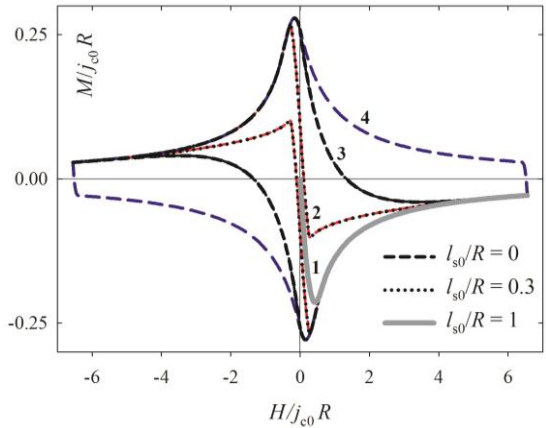
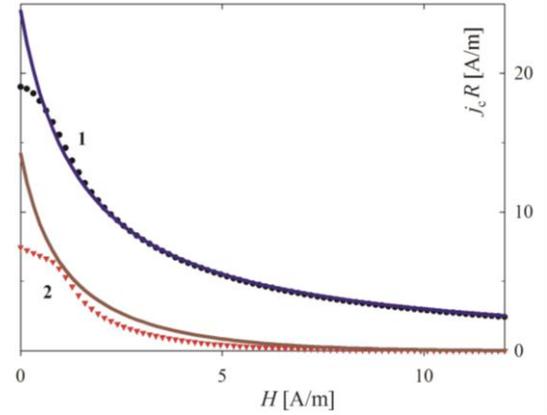

**Figure 1.** Magnetization loops $M(H)$ computed by ECSM. The curves with $H_{irr} = 10H_p$: reversible curve $l_{s0}/R = 1$ (1), $l_{s0}/R = 0.3$ (2), $l_{s0}/R = 0$ (3); the near symmetric hysteresis with $l_{s0}/R = 0$, $H_{irr} = 200H_p$ (4).

**Figure 2.** Critical current density $j_c(H)$ obtained from $M(H)$ loops. Curves (1) were obtained from the hysteresis (4) shown in figure 1 and curves (2) from the loop (2) using the Bean model (points) and ECSM (lines).

## 3. Applications

ECSM is useful for simple estimations of various parameters of superconducting samples. The model considered is applicable mainly to samples in the form of a long cylinder or a plate, for which the demagnetization factor is zero. But the penetration of the magnetic flux into samples of other shapes at high magnetic fields occurs in much the same manner [14] and the model can also be used for approximate estimations. Polycrystalline and heterogeneous superconductors are also subjects of ECSM (see subsection 3.4). The fishtail effect on magnetization loops or the peak effect [15] can also be treated by ECSM [3, 11-13]. For this purpose, it is necessary to determine the function describing the peak in the dependence $j_c(H)$ and the corresponding dip in the dependence $l_s(H)$ [3].

### 3.1. Trapped field
The computed distributions $B(r)$ gives the magnetic field in the middle of the sample and the trapped field distribution. The dependence of the magnetic field in the sample middle on the external magnetic field is presented on figure 3. The maximal value of the trapped field is equal to the full penetration field $H_p$. Increase of the ratio $l_s/R$ leads to a decrease of the trapped field in the sample middle.

The spatial distribution of the field trapped in the sample after the field cycle is drawn on figure 4. The trapped field follows a conic profile. Increase of $l_s/R$ sinks this cone (not displayed).

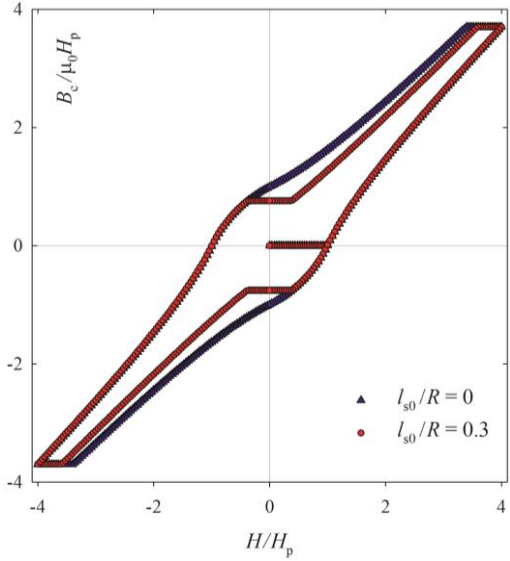
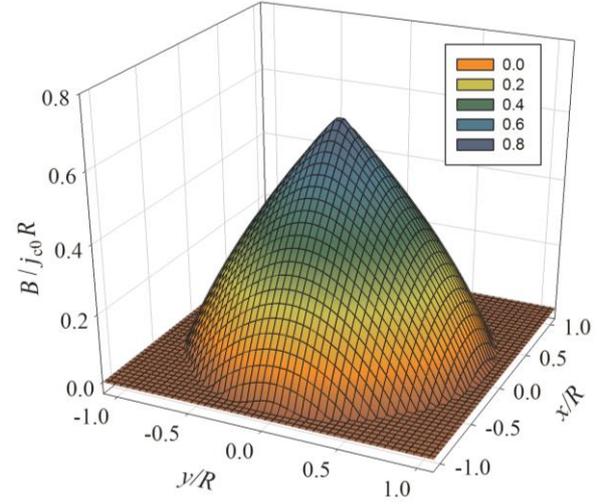

**Figure 3.** Magnetic field $B_c$ in the sample middle.

**Figure 4.** Calculations of magnetic field $B$ trapped in the sample.

*3.2. Critical current density*

The Bean model is commonly used to determine the critical current density $j_c$ and its field dependence from magnetic measurements. However it can give incorrect results in some magnetic field ranges, e.g. in the vicinity of $H = 0$. An alternative method is to fit different analytical expressions for $j_c(H)$. The relevance of the dependence $j_c(H)$ is judged from agreement between the computed and experimental magnetization loops. Simple dependencies $j_c(H) = $ const, $j_c(H) \sim 1/H$ and $j_c(H) \sim \exp(-H)$ do not lead to agreement between the calculated magnetization loop and the experimental data in strong and weak fields simultaneously. Based on many performed fittings, the dependence $j_c(H)$ was chosen, which is proportional to $1/H$ in weak fields and decreases exponentially in strong fields,

$$j_c(H) = j_{c0}\left[\left(|H|/H_1\right)^\gamma + \exp\left(|H|/H_2\right)\right]^{-1}, \qquad (2)$$

where $j_{c0}$ is the value of $j_c(H)$ for $H = 0$, $H_1$ and $H_2$ are the parameters specifying the scales of field decreasing, the index $\gamma$ is equal to 1 at low temperatures and it may be smaller than 1 at higher temperatures. Given $H_2 \sim 0.1 H_{c2}$, the calculated values of $j_c$ tend to zero for $H \geq H_{c2}$.

For near symmetric magnetization loops, the dependence $j_c(H)$ expressed by function (2) coincides over the entire field range, except for weak fields $H$, with the Bean model curve (see curves 1 on figure 2). In this case the contribution from the equilibrium magnetization of the surface layer is negligible because $l_s \ll R$. If values of $l_s$ and $R$ are comparable then the surface layer should be accounted. The surface layer of the superconducting sample with the thickness $l_s$ is not involved in supercurrent transport. The surface layer depth depends on the external magnetic field and the temperature. The simple form is used for the dependence $l_s(H)$ [4]: $l_s(H) = l_{s0} + (R - l_{s0}) H/H_{irr}$, where $l_{s0}$ is the value of $l_s$ at $H = 0$, $H_{irr}$ is the irreversibility field. When the critical current density is estimated, the area of the surface layer should be omitted such that the averaged critical current density $j_c(H)$ becomes dependent on $R$. The following dependence $j_c(H)$ is suggested

$$j_c(H) = j_{c0} \frac{\left[(1 - l_{s0}/R)(1 - H/H_{irr})\right]^n}{\left[(|H|/H_1)^\gamma + \exp(|H|/H_2)\right]}, \qquad (3)$$

here $n$ is the form-dependent power [4]: $n = 3$ for a cylindrical sample and $n = 2$ for a thin plate oriented along the field. Figure 2 demonstrates that account of $l_s$ for $l_{s0} = 0.3R$ (curve 2) is significantly modified the dependence $j_c(H)$. Difference between the curves obtained by the Bean model and the ECSM curves are clearly apparent near $H = 0$.

*3.3. Full penetration field*
In ECSM the full penetration field $H_p$ is calculated numerically and the resulted values of $H_p$ are some smaller than the Bean model values of $H_p = j_c R$. Formula (4) gives approximate values of $H_p$.

$$H_p = \frac{2H_1 H_2}{H_1 + H_2} \ln\left[1 + \frac{j_{c0} R (H_1 + H_2)}{2 H_1 H_2}\right]. \qquad (4)$$

This analytic expression is obtained from formula (2) and fitting numerically computed values of $H_p$.

*3.4. Scale of screening current*
In formulae 3 and 4 the size $R$ determines the circulation region of the screening current. For granular superconductors, there is an ambiguity in choosing the scale of $R$ [9, 16]: It can be chosen as the radius of the sample or as the average effective radius of the grains. The screening current can also circulate in clusters consisting of several grains. In order to determine the critical current density from magnetic measurements, it is necessary to establish the scale of the circulation of the screening current, because the magnetization hysteresis width is characterized by the product $j_c R$. But the circulation radius of the screening current also defines the hysteresis asymmetry by means of the relation $l_s/R$. So given $l_{s0}(T) \approx \lambda(T)$, one can find the circulation scale of the screening current from ECSM fitting of an asymmetric magnetization hysteresis [4]. This procedure is useful to accurate estimating of the intragrain critical current density. For heterogeneous samples the porosity must be also accounted for estimating of $j_c$ [2, 7]. For the granular superconductors analyzed earlier [2, 7, 9-11, 13], the effective radius satisfying the fitting is about the averaged grain size in the *ab* plane determined from scanning electron microscopy images.

## 4. Conclusion

Magnetization loops of type II superconductors are computed and parameterized by the extended critical state model. The magnetic flux distributions in the sample and the trapped field have been calculated. Asymmetry of magnetization hysteresis relative to the $M = 0$ axis originates due to the equilibrium magnetization of the surface layer. So asymmetry of the magnetization loop depends on the ratio of the surface layer depth to the radius of the screening current circulation which can be the sample size or the grain size. In the ranges of temperature and magnetic field where the depth of magnetic field penetration $\lambda$ is comparable with the radius of the screening current circulation, the critical current density and the trapped field depend on $l_s/R$ also. It is important for polycrystalline superconductors and for large grain superconductors at high magnetic fields and temperatures.